# Engine efficiency at maximum power, entropy production and equilibrium thermodynamics

Kamal Bhattacharyya[*]

Department of Chemistry, University of Calcutta, Kolkata 700 009, India

**Abstract**

The Carnot engine sets an upper limit to the efficiency of a practical heat engine. An arbitrary irreversible engine is sometimes believed to behave closely as the Curzon-Ahlborn engine. Efficiency of the latter is obtained commonly by invoking the maximum power principle in a non-equilibrium framework. We outline here two plausible routes within the domain of classical thermodynamics to arrive at the same expression. Further studies on the performances of available practical engines reveal that a simpler approximate formula works much better in respect of bounds to the efficiency. Putting an intermediate-temperature reservoir between the actual source and the sink leads to a few interesting extra observations.

* pchemkb@gmail.com; pchemkb@yahoo.com

# 1. Introduction

A Carnot engine (CE) is an example of an ideal, reversible heat engine that takes up $Q_1$ heat from a source kept at temperature $T_1$, converts a certain amount $\omega$ of it to work and rejects the rest heat $Q_2$ to a sink at temperature $T_2$. The efficiency of this engine is given by

$$\eta_{CE} = 1 - Q_2/Q_1 = 1 - T_2/T_1. \qquad (1)$$

The system undergoes a cyclic process with no change of entropy. Entropy changes of the sink and the source balance in a CE to yield the second equality in (1).

Considerable recent interest has been focused on the Curzon-Ahlborn engine (CAE) [1] that has the efficiency

$$\eta_{CAE} = 1 - \sqrt{T_2/T_1}. \qquad (2)$$

It is based on a principle called the 'efficiency at maximum power' (EMP). Expression (2) has since been a subject of fervent attention (see, e.g., [2 – 12] and references quoted therein). The initial significance [2 – 5] was quickly gained in varying circumstances. Meanwhile, a successful extension to the quantum-mechanical domain [6] followed. The practical situations were subsequently analyzed in detail [7 – 10].

A simple way to arrive at $\eta_{CAE}$ is the following. Suppose that the working substance is at a temperature $T_A$ while it takes up the heat. Similarly, during heat rejection to the sink, it is assumed to be at another fixed temperature $T_B$. Then the rates of heat flow and the power $P$ are represented by

$$\dot{q}_1 = K_1(T_1 - T_A), \quad \dot{q}_2 = K_2(T_B - T_2); \\ P = \dot{q}_1 - \dot{q}_2, \qquad (3)$$

with constants $K_1$ and $K_2$. Using the entropy balance equation for the working fluid in the form

$$\dot{q}_1/T_A = \dot{q}_2/T_B \qquad (4)$$

one can express *P* in terms of a single variable $T_A$ (or $T_B$) and maximize it to finally find the optimum values as

$$T_A(\text{opt}) = \left(K_1 T_1 + K_2 \sqrt{T_1 T_2}\right) \Big/ (K_1 + K_2),$$
$$T_B(\text{opt}) = \left(K_2 T_2 + K_1 \sqrt{T_1 T_2}\right) \Big/ (K_1 + K_2). \tag{5}$$

Defining $\eta$ by

$$\eta = 1 - (\dot{q}_2 / \dot{q}_1), \tag{6}$$

and employing the optimum values (5) in (3), expression (2) for the EMP is obtained from (6).

The kind of finite-time thermodynamic approach as above has, however, confronted severe criticisms [11]. It is also not clear whether the alleged success of CAE in reproducing efficiencies of practical engines is fortuitous [4, 7, 12]. Indeed, with data for 10 practical engine efficiencies [13], an early comparative survey [12] revealed no great promise in (2). However, a resurgence in this area of EMP occurred after an ingenious route was provided [14] via Onsager's formulation of linear irreversible thermodynamics. Varying types of extension along this line has since then continued [15 - 28]. A novel stochastic approach was pursued [15, 20, 21]; finite-time CE was studied [16, 17, 28] within the linear regime; a sort of universality of the CAE has later been also justified [18, 19, 22, 26] *beyond* the linear regime. Particularly important here is the emergence of bounds [22, 26] to (2). These were tested [22] with 13 realistic cases [13, 29]. However, once again, the worth is not immediately apparent. Gradually, therefore, a belief has again started to develop that $\eta_{\text{CAE}}$ does not reflect the true state of affairs [23 - 25]. Attempt has also been made [27] to express the EMP as a nonlinear function of the difference of chemical potentials between the materials of the reservoirs for a chemical engine.

In view of the above remarks, here we address primarily three issues. First, we put forward two strictly equilibrium thermodynamic avenues to arrive at (2). Our second endeavor will be to scrutinize the adequacy of (2) vs. a simpler estimate in predicting the efficiencies of randomly selected working engines. As we shall see, this latter approach is more economical and faithful in respect of bounds. In fine, we observe how the inclusion of a third reservoir at some intermediate temperature leads to certain interesting conclusions.

## 2. Thermodynamic routes

In conventional thermodynamics, we do not talk of power. So, EMP cannot be the guiding principle here. Instead, we provide two strictly thermodynamic arguments to arrive at (2). These include (a) the use of finite reservoirs and (b) a specific choice of an average irreversibility indicator for infinite reservoirs.

### A. Finite reservoirs with reversible heat transfers

Suppose that the source and the sink have the same *finite* heat capacity $C$. Then, after each cycle, source temperature will decrease and sink temperature will increase to reach a common temperature $T'$ at the end, and the cycle will stop. Assume reversible heat transfers, for simplicity, to obtain $T'$. Results are the following:

$$\Delta S_{SURR} = \int_{T_1}^{T'} CdT/T + \int_{T_2}^{T'} CdT/T = 0;$$
$$T' = \sqrt{T_1 T_2}. \tag{7}$$

The overall efficiency may now be calculated as follows, yielding (2) again:

$$\omega = C(T_1 - T') - C(T' - T_2) = C(T_1 + T_2 - 2T') = C(\sqrt{T_1} - \sqrt{T_2})^2;$$
$$q_1 = C(T_1 - T') = C\sqrt{T_1}(\sqrt{T_1} - \sqrt{T_2}); \tag{8}$$
$$\eta = \omega/q_1 = 1 - \sqrt{T_2/T_1}.$$

It may be appealing to extend the above logic to a case with different heat capacities of the source and the sink ($C_1$ and $C_2$, respectively). We summarize the final findings, for convenience:

$$T' = T_1^x T_2^{1-x}; \quad x = C_1/(C_1+C_2);$$
$$\eta = 1 - \frac{(T'-T_2)C_2}{(T_1-T')C_1} = 1 - \frac{(1-x)(T_1^x T_2^{1-x} - T_2)}{x(T_1 - T_1^x T_2^{1-x})}. \quad (9)$$

Note that the outcomes are much more complicated and, importantly, the efficiency is dependent on the 'heat capacity fraction' now. However, we shall not henceforth consider the case of finite reservoirs in any discussion.

**B. Infinite reservoirs with arbitrary irreversible engines**

We now consider standard infinite reservoirs. Take an arbitrary irreversible engine that takes up $q_1 = Q_1$ heat from the source at $T_1$ and delivers $q_2 = \beta Q_2$ heat to the sink at $T_2$ where $Q_1$ and $Q_2$ are defined in (1). The corresponding efficiency reads as

$$\eta = 1 - \frac{q_2}{q_1} = 1 - \frac{\beta Q_2}{Q_1} = 1 - \frac{\beta T_2}{T_1} = 1 - \frac{\beta}{\lambda}. \quad (10)$$

where $\lambda = T_1/T_2$. Here, $\beta$ is some kind of a 'irreversibility index' that is bounded by $1 \leq \beta \leq \lambda$. It grows from unity to indicate enhanced irreversible character of an engine.

An arbitrary engine can have any possible value of $\beta$ between 1 and $\lambda$. Therefore, we have three simple alternatives for the 'average' $\beta$:

$$\beta(\text{AM}) = \frac{1+\lambda}{2} \geq \beta(\text{GM}) = \sqrt{\lambda} \geq \beta(\text{HM}) = \frac{2\lambda}{1+\lambda}. \quad (11)$$

These values respectively signify the arithmetic mean (AM), the geometric mean (GM) and the the harmonic mean (HM). From (10), one thus obtains three average estimates for $\eta$ that are ordered as

$$\eta(\text{AM}) = \frac{\eta_{\text{CE}}}{2} = \eta_{\text{L1}} \leq \eta(\text{GM}) = 1 - \sqrt{1-\eta_{\text{CE}}} \leq \eta(\text{HM}) = \frac{\eta_{\text{CE}}}{2-\eta_{\text{CE}}} = \eta_{\text{U1}}. \quad (12)$$

It is easy to identify $\eta(\text{GM})$ with $\eta_{\text{CAE}}$, and then to notice that the inequalities in (12) are known [22, 26]. Indeed, they define the primary lower and upper bounds, and hence denoted respectively by $\eta_{L1}$ and $\eta_{U1}$; but, one can cook up more of such inequalities, to be outlined later.

## 3. Entropy production, scaling and balance

Let us now look for the entropy production (EP), given by

$$\Delta S = \frac{q_2}{T_2} - \frac{q_1}{T_1} = \frac{Q_1}{T_1}(\beta - 1), \qquad (13)$$

It is expedient now to define a dimensionless EP. To this end, we notice that $\Delta S$ in (13) reaches its maximum value $\Delta S_m$ when the condition $\eta = 0$, i.e., $Q_1 = \beta Q_2$, or $\beta = \lambda$, holds. The dimensionless (intensive) EP is now designated as

$$\overline{\Delta S} = \frac{\Delta S}{\Delta S_m} = \frac{\beta - 1}{\lambda - 1}. \qquad (14)$$

Note that $\overline{\Delta S}$ refers to the probability that heat will be unavailable for work; it varies within zero and unity. To confer the same kind of symmetry in the efficiency part, we construct also

$$\overline{\eta} = \frac{\eta}{\eta_{CE}} = \frac{\lambda - \beta}{\lambda - 1}. \qquad (15)$$

It defines the probability of conversion of heat to work and obeys $0 \leq \overline{\eta} \leq 1$. Thus, there is a conservation law in the form

$$\overline{\eta} + \overline{\Delta S} = 1 \qquad (16)$$

Let us now emphasize that, generally there is a competition between the *natural* tendency to increase $\overline{\Delta S}$ and *human* endeavor to increase $\overline{\eta}$. One should therefore be interested to find an optimal situation with respect to $\beta$ in any practical case. Notably, only when $\beta = \beta(AM)$, both $\overline{\Delta S}$ and $\overline{\eta}$ have the *same* optimum value. Hence, they balance each other exactly at this point. Therefore, $\eta = \eta(AM)$ possesses an edge over $\eta(GM) = \eta_{CAE}$ for which $\overline{\eta} > \frac{1}{2}$.

## 4. Improved bounds

Known upper and lower bounds to (2), respectively given by $\eta_{U1}$ and $\eta_{L1}$, are already found in (12). General results in this context might be summarized via $\eta$ in (10) as

$$\eta(\beta > \sqrt{\lambda}) < \eta_{CAE}; \eta(\beta < \sqrt{\lambda}) > \eta_{CAE}. \qquad (17)$$

Thus, employing $\beta = \lambda^{2/3}$, we obtain a lower bound $\eta_{L2}$. Similarly, the choice $\beta = \lambda^{1/3}$ leads to another upper bound $\eta_{U2}$. These bounds are given by

$$\eta_{CAE} \geq 1 - (1-\eta_{CE})^{1/3} = \eta_{L2}; \eta_{CAE} \leq 1 - (1-\eta_{CE})^{2/3} = \eta_{U2}. \qquad (18)$$

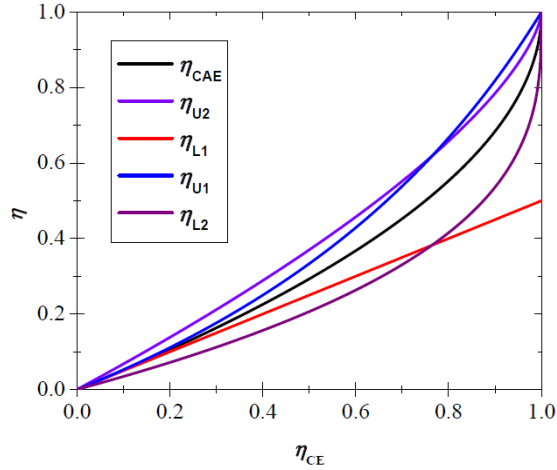

**Figure 1.** Performance of a few upper and lower bounds (see text) to $\eta_{CAE}$.

They are seen to be more efficient in regions where $\eta_{CE}$ is high. Figure 1 shows the efficacy of such bounds. However, search of bounds may continue endlessly. For example, by choosing $\lambda^{2/3}$ and $\lambda^{1/3}$ and taking the AM, one can find a very tight lower bound as

$$\eta_{CAE} \geq 1 - [(1-\eta_{CE})^{1/3} + (1-\eta_{CE})^{2/3}]/2 = \eta_{L3}. \qquad (19)$$

## 5. Case-studies and further remarks

We mentioned before that the CAE efficiency is not unquestionably believed [4, 7, 12] to correspond to efficiencies of realistic engines. Later, study of a Brownian engine [15] also revealed that the EMP scheme does not lead to (2). On the other hand, a molecular dynamics simulation [16] found expression (2) only under a limiting situation. So, one feels obliged to put (2) to test.

## A. Demonstrative performance

Perusal of the relevant literature reveals that test cases [22] show no great promise either, even if we provide allowance for the bounds [22, 26] to (2). A glimpse at Table 1 will make the point quite clear. Here, the plants are ordered as they appeared in [22].

Table 1. Efficiencies of some working engines [13, 22, 29] and theoretical estimates. Bounds $\eta_{L1}$ and $\eta_{U1}$ are defined in Eq. (12).

| Plant | $T_1, T_2$ | $\eta$(obs) | $\eta_{CE}$ | $\eta_{CAE}$ | $\eta_{U1}$ | $\eta_{L1}$ |
|---|---|---|---|---|---|---|
| 1  | 566, 283 | 0.35 | 0.50 | 0.29 | 0.33 | 0.25 |
| 2  | 601, 290 | 0.34 | 0.52 | 0.30 | 0.35 | 0.26 |
| 3  | 581, 288 | 0.36 | 0.50 | 0.30 | 0.34 | 0.25 |
| 4  | 562, 289 | 0.34 | 0.49 | 0.28 | 0.32 | 0.24 |
| 5  | 727, 288 | 0.40 | 0.60 | 0.37 | 0.43 | 0.30 |
| 6  | 838, 298 | 0.36 | 0.64 | 0.40 | 0.48 | 0.32 |
| 7  | 573, 298 | 0.30 | 0.48 | 0.28 | 0.32 | 0.24 |
| 8  | 523, 353 | 0.16 | 0.33 | 0.18 | 0.19 | 0.16 |
| 9  | 583, 298 | 0.19 | 0.49 | 0.29 | 0.32 | 0.24 |
| 10 | 783, 298 | 0.34 | 0.62 | 0.38 | 0.45 | 0.31 |
| 11 | 698, 298 | 0.28 | 0.57 | 0.35 | 0.40 | 0.29 |
| 12 | 963, 298 | 0.32 | 0.69 | 0.44 | 0.53 | 0.35 |
| 13 | 953, 298 | 0.34 | 0.69 | 0.44 | 0.52 | 0.34 |

Let us focus attention on two specific issues. First, the upper ($\eta_{U1}$) and lower ($\eta_{L1}$) bounds to $\eta_{CAE}$ are not tight enough, particularly when the efficiency itself is high. Indeed, certain cases show as large as 20% departures. Secondly, it is quite frustrating to notice that the observed values [$\eta$(obs)] are *greater* than $\eta_{U1}$ for plants 1, 3 and 4. Conversely, cases 9, 11 and 12 take on values *less* than what $\eta_{L1}$ suggest. Plants 8 and 13 show *just* a borderline behavior in respect of $\eta_{L1}$.

The above observations again suggest that efficiencies of practical engines do not obey the quoted bounds to $\eta_{CAE}$. One early opinion was that, real engines do not satisfy the EMP principle [22, 24]. Some other works [23, 25] suggested also that $\eta_{CAE}$ does not *always* follow from the above principle, even in the linear-response regime [25]. Thus, a fresh look at the problem is obligatory. The task is a clean rationalization of the data presented in Table 1.

**B. A random selection approach**

A naive but straightforward approach could be the following. Given a fixed $\lambda$, the upper limit to efficiency ($\eta_{CE}$) is naturally fixed. We know that the efficiency of any arbitrary engine may have any value between 0 and $\eta_{CE}$. Assuming *equal a priori* probability of any such value, in lieu of deliberately introducing an additional principle or constraint, one obtains as average ($\eta_A$) and standard deviation ($\Delta\eta$) the estimates

$$\eta_A = \eta_{CE}/2; \quad \Delta\eta = \eta_{CE}/\sqrt{12}. \tag{20}$$

The first result is a direct AM as well. It appeared earlier [13, 14, 22], and also in (12) as a lower bound to $\eta_{CAE}$ [i.e., $\eta_{L1} = \eta_A = \eta(AM)$]. From the second result in (20), we have bounds to $\eta$ as

$$\eta_{CE}/2 - \eta_{CE}/\sqrt{12} \leq \eta_A \leq \eta_{CE}/2 + \eta_{CE}/\sqrt{12}. \tag{21}$$

Though the $\eta_A$ values are sometimes not quite close to $\eta(\text{obs})$, we checked that no violation of (21) has occurred in any case. Advantage of this simplistic argument is thus obvious.

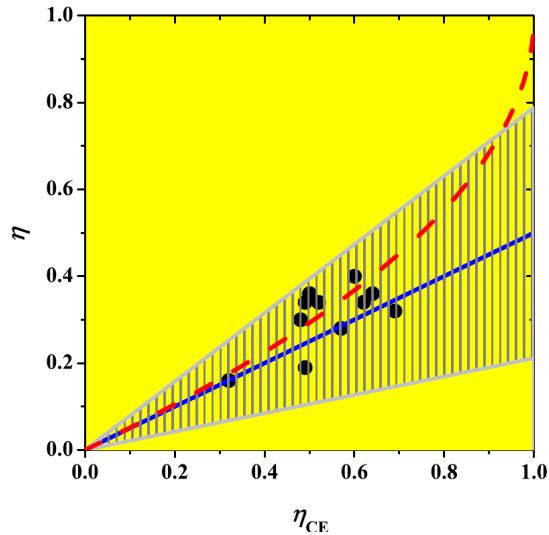

**Figure 2. Plots of average $\eta$ (middle solid line, $\eta_A$), its bounding regions [eq. (21): extreme solid lines], practical cases (points) and $\eta_{CAE}$ (red dashed line) as functions of $\eta_{CE}$.**

Figure 2 testifies the bounding property of (21) beyond doubt. One may also notice that both $\eta_{U1}$ and $\eta_{CAE}$ approach $\eta_{CE}$ as $\eta_{CE} \rightarrow 1$. In our case, however, neither the upper bound nor the average would ever reach $\eta_{CE}$. This is another desirable feature from a pragmatic standpoint. Beyond $\eta_{CE} \approx 0.6$, the $\eta$-values tend more towards $\eta_A$ than $\eta_{CAE}$. Therefore, it is unlikely that an arbitrary engine will have $\eta$ closer to $\eta_{CAE}$, when $\eta_{CE}$ itself is large. Indeed, unless special conditions are imposed during construction, any such observation that $\eta \approx \eta_{CAE}$ (and not $\eta \approx \eta_A$) is liable to be singular, especially at large $\eta_{CE}$.

**C. Role of an intermediate-temperature reservoir**

It is a common knowledge that one does not gain by putting an intermediate-temperature reservoir between the actual source and the sink only in case of a CE. Suppose that the intermediate temperature is $T_j$ where $T_2 < T_j < T_1$. We then have the following results for work, respectively in case of a single engine ($\omega$) and a composite set up using that engine twice ($\omega_C$):

$$\omega = Q_1 \eta; \quad \omega_C = Q_1 \eta_1 + Q_1 (1-\eta_1)\eta_2. \tag{22}$$

The gain in terms of work thus turns out to be

$$\Delta \omega = Q_1 [\eta_1 + (1-\eta_1)\eta_2 - \eta]. \tag{23}$$

For convenience, we also define $\lambda_1 = T_1/T_j$, $\lambda_2 = T_j/T_2$. Let us now choose efficiencies in the form (10). This means, $\eta_k = 1 - \beta_k/\lambda_k$ (k = 1, 2) and $\eta = 1 - \beta/\lambda$. From (23), in such a situation, we find that

$$\Delta \omega = Q_1 [(\beta/\lambda) - (\beta_1 \beta_2 / \lambda_1 \lambda_2)] = (Q_1/\lambda)(\beta - \beta_1 \beta_2). \tag{24}$$

The following observations are relevant: (i) For the CE case, $\beta_1 = \beta_2 = \beta = 1$. So, we obtain here $\Delta \omega = 0$. (ii) A specific irreversible engine may be characterized by the same ratio of $\beta_k/\lambda_k = \beta/\lambda = \phi$. This would imply the same degree of irreversibility or same efficiency. Then (24) yields

$$\Delta \omega = Q_1 \phi (1-\phi) \geq 0 \tag{25}$$

since $\phi \leq 1$. (iii) Another kind of choice for an irreversible engine might be $\eta = \alpha \eta_{CE}$, with $0 < \alpha$

< 1. This means, such an engine always offers a specific fraction of the Carnot efficiency. Putting it in (23), one finds again

$$\Delta\omega = \alpha Q_1 (1-\alpha) \frac{(T_1 - T_j)(T_j - T_2)}{T_1 T_j} > 0. \tag{26}$$

(iv) Whereas the above two general types of irreversible engines yield $\Delta\omega > 0$, and the same is true also of $\eta$(AM) for which $\alpha = \frac{1}{2}$, the inequality $\Delta\omega > 0$ is not satisfied by $\eta_{CAE}$. Like the CE, (24) shows $\Delta\omega = 0$ for the CAE because here $\beta = \beta_1\beta_2$. (v) Indeed, the result $\Delta\omega = 0$ follows for any engine whose efficiency can be expressed in the form

$$\eta = 1 - (T_2 / T_1)^\gamma; \; 0 < \gamma < 1. \tag{27}$$

There is, however, no simple way to arrive at the special form (27). Interesting at this point is to also note that $\eta$(HM) in (12) yields $\Delta\omega < 0$. The behaviors in terms of (24) are shown in Figure 3 as a function of $T_j$. We note, whereas $\eta$(HM) and $\eta$(GM) do not violate any general principle, they do perform rather unusually in respect of extracting advantages via insertion of the extra reservoir. The HM form for $\eta$ is always a disadvantage, and this is why we have not put any emphasis on $\eta$(HM) in our discussion. It attains its minimum at $T_j = (T_1 T_2)^{1/2}$, while the same point refers to the maximum advantage for $\eta$(AM).

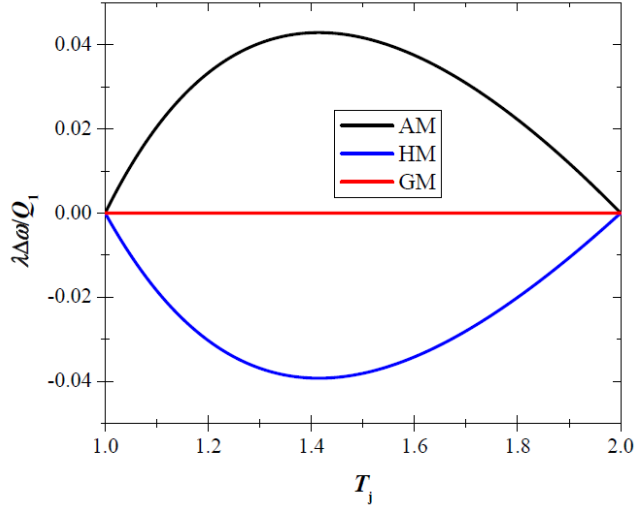

**Figure 3.** Plots of $\lambda\Delta\omega/Q_1$ for the three different cases [see (12) and (24)] of average $\eta$ at different intermediate temperatures with $1 = T_1 < T_j < T_2 = 2$.

**D. Special role of the sink**

In discussions of work and entropy production, the sink plays a special role. For example, one finds from (13) that

$$T_2 \Delta S_m = Q_1 \eta_{CE} = \omega_m \tag{28}$$

where the subscript '$m$' refers to the maximum value. Coupled with (16), it gives

$$T_2 \Delta S + \omega = \omega_m. \tag{29}$$

The maximum possible work is thus split into an available part and an unavailable part, but $T_1$ never shows up in either (28) or (29). There is thus a strong hint that setting $T_2 = 0$ would be problematic. If we now insert a third intermediate-temperature reservoir and employ two engines, the difference quantities will obey

$$T_2 \Delta(\Delta S) + \Delta \omega = 0. \tag{30}$$

In other words, the CAE will show no change in $\Delta S$ too if the composite set up is run.

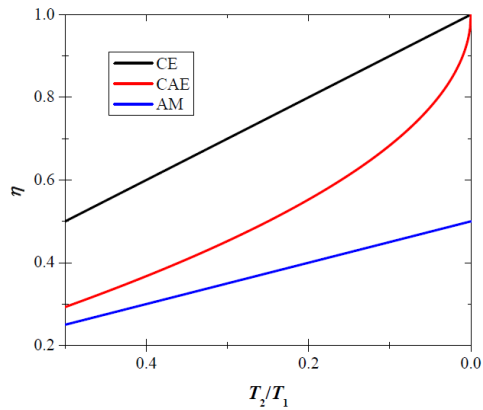

**Figure 4.** Plot of $\eta$ vs. $T_2/T_1$ starting from a value of ½ for the three cases of concern. The slope for the CAE is notable as $T_2$ gradually diminishes.

Sink shares another important feature. It is well-known that the sensitivity of $\eta_{CE}$ is more with respect to slight changes of $T_2$ than $T_1$. But, this sensitivity with $T_2$ increases without bound for $\eta_{CAE}$ as $T_2 \to 0$. Figure 4 shows how quickly the slope of $\eta_{CAE}$ increases as $T_2$ decreases towards zero for some fixed $T_1$. Once again, no violation of any general principle is apparent, but the observation is somewhat disturbing. Note that $\eta_A$ behaves quite normally in this respect, like $\eta_{CE}$.

### E. More intermediate-temperature reservoirs

An interesting special situation arises when we put ($N$-1) intermediate-temperature reservoirs between $T_1$ and $T_2$ as $T_2(1) > T_2(2) > \ldots > T_2(N-1) > T_2 = T_2(N)$ and run $N$ engines of the same nature. The net $\Delta S$ would be

$$\Delta S = Q_1 \prod_{k=1}^{N}(1-\eta_k)/T_2 - Q_1/T_1. \tag{31}$$

This can be rearranged, by defining $\eta_k = (1 - \beta_k/\lambda_k)$, $\lambda_k = T_2(k-1)/T_2(k)$ as

$$T_2 \Delta S = Q_1 \left[\prod_{k=1}^{N}(\beta_k/\lambda_k) - 1/\lambda\right]; 1 < \beta_k < \lambda_k. \tag{32}$$

For any irreversible engine, it may seem at first sight that, when $N \to \infty$, the first factor in the brackets would vanish because more and more smaller fractions are gradually getting multiplied. But, this is not so. This counter-intuitive result has its origin in thermodynamics. The second law restricts that $\Delta S < 0$ will not be allowed. The mathematical logic here is, $\Pi \lambda_k = \lambda$ by choice, and since all $\beta_k$ terms are greater than unity, the first factor will always exceed the second. Only for the CE we shall have $\Delta S = 0$, as here all $\beta_k = 1$.

### 6. Conclusion

In summary, we have outlined two ways of arriving at the CAE efficiency without invoking the EMP. During the exploration, we have noted certain inequalities. Some of these are found to be useful later in providing better bounds to $\eta_{CAE}$. Efficiencies of most practical engines, however, lie outside the normal bounds to $\eta_{CAE}$. As a better alternative, we have found that $\eta_A$ serves the purpose in a cogent manner [see eq. (21) and Figure 2] in respect of bounds. Finally, we study the effect of inserting a third, intermediate-temperature reservoir. The CAE is seen to offer absolutely no advantage; it resembles the ideal CE in this regard. Moreover, the rate of change of $\eta_{CAE}$ with $T_2$ (at fixed $T_1$) as $T_2 \to 0$ is very singular (see Figure 4), unlike the CE case. Both these features confer $\eta_A$ some edge over $\eta_{CAE}$ and they emerge in the latter case because of the peculiar expression (2), being a part of the specific family (27). Indeed, these observations led us to study the behavior of practical engines against $\eta_A$ and its bounds in Sec. 5. Keeping in mind recent interests on the CAE [30-34] and its historical traces dating back to 1929 [33], the present endeavor may be of value. In particular, we note that the discussion in Sec. 2a has been covered elsewhere [34], but our approach in Sec. 2b is very different. The effect of insertion of an intermediate-temperature reservoir in this context is new. Also new is our observation on the increase of $\eta_{CAE}$ without limit as $T_2 \to 0$. It may serve as a challenge to experimentalists whether such a theoretical prediction may be tested at all keeping the sink temperature very low.